\documentclass[doublecol,figures]{epl2} 
\usepackage{amsmath}
\usepackage{amssymb}

\makeatletter
\@ifundefined{textcolor}{}
{
 \definecolor{BLACK}{gray}{0}
 \definecolor{WHITE}{gray}{1}
 \definecolor{RED}{rgb}{1,0,0}
 \definecolor{GREEN}{rgb}{0,1,0}
 \definecolor{BLUE}{rgb}{0,0,1}
 \definecolor{CYAN}{cmyk}{1,0,0,0}
 \definecolor{MAGENTA}{cmyk}{0,1,0,0}
 \definecolor{YELLOW}{cmyk}{0,0,1,0}
}

\makeatother

\begin{document}

\title{Emergence of a non trivial fluctuating phase in the XY{-rotors} model on regular networks}

\author{Sarah De Nigris \and Xavier Leoncini}

\institute{Centre de Physique Th\'{e}orique
, CNRS - Aix-Marseille Universit\'{e}, Luminy, Case 907, F-13288 Marseille
cedex 9, France}

\abstract{We study  {an XY-rotor model on regular one dimensional lattices by varying the 
number of neighbours. The  parameter $2\ge\gamma\ge1$ is defined. $\gamma=2$ corresponds
to mean field and $\gamma=1$ to nearest neighbours coupling}.  We find that for $\gamma<1.5$
the system does not exhibit a phase transition, while for $\gamma>1.5$
{the mean field second order transition is recovered}.
{ For the critical value $\gamma=\gamma_c=1.5$, {the systems can be in a non trivial fluctuating phase for which} the magnetisation 
shows important fluctuations in a given temperature range, {implying} an infinite susceptibilty. {For all values of $\gamma$} the magnetisation is computed analytically in the low temperatures
range and the magnetised versus non-magnetised state which depends on the value of $\gamma$ is recovered,}
confirming the critical value $\gamma_{c}=1.5$.}

\pacs{05.20.-y}{Classical statistical mechanics}
\pacs{05.45.-a}{Nonlinear dynamics and chaos}

\maketitle
In {the last two decades}, systems with long-range interactions have attracted
increasing attention and have been widely studied \cite{Dauxois_book2002,Campa09}.
{In systems with short range interactions, the  property of \emph{additivity} allows to construct the canonical ensemble from the microcanonical,
the two approaches being equivalent in the thermodynamic limit \cite{galla}. In constrast, the lack of additivity {adds another layer of complexity to} the picture when dealing 
with systems interacting via a long-range  potential \cite{inequivalence,ens_inequivalence,Torcini99,Pluchino2007,Chavanis2006,Latora2002}, giving for instance rise to possible negative specific heat in the microcanonical ensemble. Another peculiar feature corresponds to the fact that
some long-range systems  may dynamically} keep track of their initial configuration, leading to long-lasting
quasistationary states (QSSs). {Peculiar of those states is their duration, which diverges with the system size \cite{Ettoumi2011}, leading to ergodicity
breaking \cite{long_relax,Campa09,chavanis2005,Latora2002}.}
 {Over the years the mean field rotator model (HMF), which corresponds to a mean field $XY$ model with an added kinetic term \cite{Antoni95}, 
has become a paradigmatic model for the study of long range systems. 
In this frame, QSSs have been extensively studied and an out of equilibrium phase transition has been displayed \cite{Chavanis-RuffoCCT07, Firpo2009}. 
Moreover, these stationary states have been shown to display intriguing regular microscopic dynamics \cite{Bachelard08, Leoncini09b} and} an oscillating metastable state was observed \cite{oscillation_hamilt},
enriching the already various scenario of the HMF model.{ Moving one step further, a coupling constant depending on the distance $r$ like $1/r^{\alpha},0<\alpha<2$ 
was introduced, giving birth to the so called $\alpha-HMF$ model \cite{Anteneodo98, Tamarit00,Giansanti02,Vandenberg10}. The parameter $\alpha$ allowed to explore the transition between 
the non-additive regime, for $\alpha<1$,  and the additive one for $\alpha>1$: the first case, belonging to the aforementioned class of long-ranged systems, 
unveiled the same degree of complexity than the HMF model, displaying as well QSSs and negative specific heat \cite{Campa02}. Relaxing the assumption of global 
coupling, the XY model with just nearest neighbours interactions has been in his turn a very fertile subject for decades of numerical studies \cite{Landau1984,Loft87,Kim94,Janke91,McCarthy86,Jain1986,Bramwell2001} . Among countless other remarkable features, this model in two dimensions shows a 
\emph{infinite order} phase transition, retrieved by Kosterlitz and Thouless \cite{Kosterlitz}, affecting the correlation function: for low temperatures it shows a power 
law decay, while it switches to an exponential behaviour for high temperatures. More recently, another issue challenged the study of long-range systems: their interplay 
with complex network topologies inspired by real world ones \cite{Barrat_book}. Concerning the XY model, we acknowledge for instance studies on random networks 
\cite{diluted_networks} or on Small-World networks \cite{XY_in_SWnet,dynamical_critical_behaviour}. }

In this Letter, we address this issue of complex networks too, investigating the transition from
short-range to long-range regime from a quite different point of view
than previous works. We chose as control parameter a \emph{topological} condition, which
is imposing the connectivity per interacting unit. We used the paradigmatic
$1$D-$XY$ model for rotors and we will show that we can identify
two limit regimes: a short-ranged one for low connectivity while,
in the limit of high connectivity, the system shows global coherence
via a second order phase transition. The main result of the paper
is, however, the emergence of a peculiar new state in between in which
the order parameter is affected by important fluctuations. Furthermore,
we will show analytically that this state stems from the special \emph{topological}
condition on the connectivity we imposed. 

In general the $XY$ model describes a system of $N$ pairwise interacting
units. At each unit $i$ is assigned a real number $\theta_{i}$ ,
which we refer to as the \emph{spin} $i$. In the following, we will
consider the $XY$ model from the point of view of classical Hamiltonian
dynamical systems by adding a kinetic energy term to the $XY$ Hamiltonian.
The total Hamiltonian \emph{$H$ }takes the form: 
\begin{equation}
H=\sum_{i=1}^{N}\frac{p_{i}^{2}}{2}+\frac{J}{2k}\sum_{i,j=1}^{N}\epsilon_{i,j}(1-\cos(\theta_{i}-\theta_{j})).\label{eq: hamiltonian}
\end{equation}
We associate to each spin $i$ a canonical momentum $p_{i}$ whose
coupled dynamics with the $\{\theta_{i}\}$ will be given by the set
of Hamilton equations:
\begin{equation}
\dot{\theta}_{i}=p_{i},\hspace{1em}\dot{p}_{i}=-\frac{J}{k}\sum_{i,j=1}^{N}\epsilon_{i,j}(\cos\theta_{j}\sin\theta_{i}-\sin\theta_{j}\cos\theta_{i}).\label{eq:dynamics-1-1}
\end{equation}
The coupling constant $J$ in Eqs. (\ref{eq: hamiltonian}) and (\ref{eq:dynamics-1-1})
is chosen positive in order to obtain a ferromagnetic behaviour and
in the following it will be set at $1$ without loss of generality.
We encode the information about the links connecting the units in
the \emph{adjacency matrix} $\epsilon_{i,j}$ :

\begin{equation}
\epsilon_{i,j}=\begin{cases}
1 & if\,\, i,j\,\, are\,\, connected\\
0 & otherwise
\end{cases}\:.\label{eq:adjacency matrix}
\end{equation}
By construction, the adjacency matrix is a symmetric matrix with null
trace. In Eq. (\ref{eq: hamiltonian}) the normalisation constant
$k$ ensures the extensivity of the energy, according to the
Kac prescription, and it corresponds to the {average} number of links per spin, {often referred to as} the \emph{average degree} of the network.
In this letter we {focus on} regular {one dimensional {rings} in which each spin 
is connected to $k/2$ neighbours on each side and we shall tune the width of this neighbourhood by adjusting $k$. 
Following the philosophy devised in \cite{diluted_networks} for random networks, instead of considering $k$ per se, we use the  parameter $\gamma$ defined by:} 

\begin{equation}
k\equiv\frac{1}{N}\sum_{i>j}\epsilon_{i,j}=\frac{2^{2-\gamma}(N-1)^{\gamma}}{N}\:,\label{eq:degree}
\end{equation}
{where  $\gamma\in[1,2]$.{ In order to get a natural number we take the integer part of Eq. (\ref{eq:degree}) once the size $N$ and $\gamma$ are fixed}. 
Given Eq. (\ref{eq:degree}) we have:
\begin{equation}
 \gamma=\frac{\log(Nk/4)}{\log((N-1)/2)}.\label{eq:gamma}
\end{equation}
 $\gamma$ offers a simple manner to shift continuously from
the short-range to the long-range regime: the case $\gamma=1$ corresponds
to the linear chain with only the two nearest neighbours coupling and, on
the other hand, $\gamma=2$ corresponds to the full {mean field} coupling of all
the spins. In the latter case the Hamiltonian in Eq. (\ref{eq: hamiltonian})
reduces to the \emph{HMF} model \cite{Antoni95}. {Hence the action of lowering $\gamma$ corresponds to a dilution
 of the number of links with the HMF as a reference.} To investigate the macroscopic behaviour of
the system, we define the magnetisation $\mathbf{M}=(m_{x},m_{y})$,
where $m_{x}=N^{-1}\sum_{i}\cos(\theta_{i})$ and $m_{y}=N^{-1}\sum_{i}\sin(\theta_{i})$.
The modulus $M=|\mathbf{M}|$ indicates the degree of coherence  of
the spin angular distribution: the incoherent state will have $M=0$,
while finite values of $M$ are naturally associated to more coherent states. Having set the structure of the lattice via the {parameter $\gamma$}, we performed
simulations in the microcanonical ensemble and we studied the evolution
of the total equilibrium magnetisation $\overline{M}$ where the bar
denotes the time average (we assume ergodicity). The system possesses
two constants of motion preserved by the dynamics: the energy $H=E$
and the total angular momentum $P=\sum_{i}p_{i}$ which are set by
the initial conditions. We chose to start the system with a Gaussian
distribution for both the spins and the momenta. We also impose
$P=0$ {which, given the equations of motion, implies as well the conservation of $Q=\sum_{i}\theta_{i}$.} The numerical integration of Eqs. (\ref{eq:dynamics-1-1})
is performed {with} the fifth optimal symplectic integrator {proposed in} \cite{McLachlan92}. {In our simulations, we chose a $\Delta t=0.05$ and 
we monitored the conservations of $E$ and $P$ to ensure the correcteness of the numerical integration}. The thermodynamic quantities are calculated by averaging over time. 
The energy density is measured as $\epsilon=E/N$ and, for the temperature, we consider its kinetic definition as the average kinetic energy per particle, 
since the average momentum is conserved and by our choice fixed to be zero. We first{ focused on the interval $\gamma<1.5$} . For
this regime, the system doesn't show a phase transition
of the order parameter.  {In fact,  for $\gamma$ approaching 1}, the system is more or less identical to a short-range  system and in that case, the Mermin-Wagner theorem
{imposes the order parameter to vanish}. Still
finite size effects are at play and the results  displayed {in Figs.~\ref{Flo:magnetisation 1.25}a and~\ref{Flo:magnetisation 1.25}b} show that 
the magnetisation {appears to decrease} with the system size {at every density energy  $\epsilon=E/N$ in the physical range}, so that in the thermodynamic
limit we expect the residual magnetisation to be zero. Nevertheless,
quasi-long-range order could still arise at finite temperatures like
in the $2-$D short-ranged $XY$-model which displays the Berezinskij-Kosterlitz-Thouless
phase transition \cite{berezinskij,Kosterlitz}. This particular phase
transition is characterized by the change in behaviour of the correlation
function, which decays as a power law at low temperatures and exponentially
in the high temperature phase. Hence 
\begin{figure}

\begin{centering}
(a)\includegraphics[width=8cm]{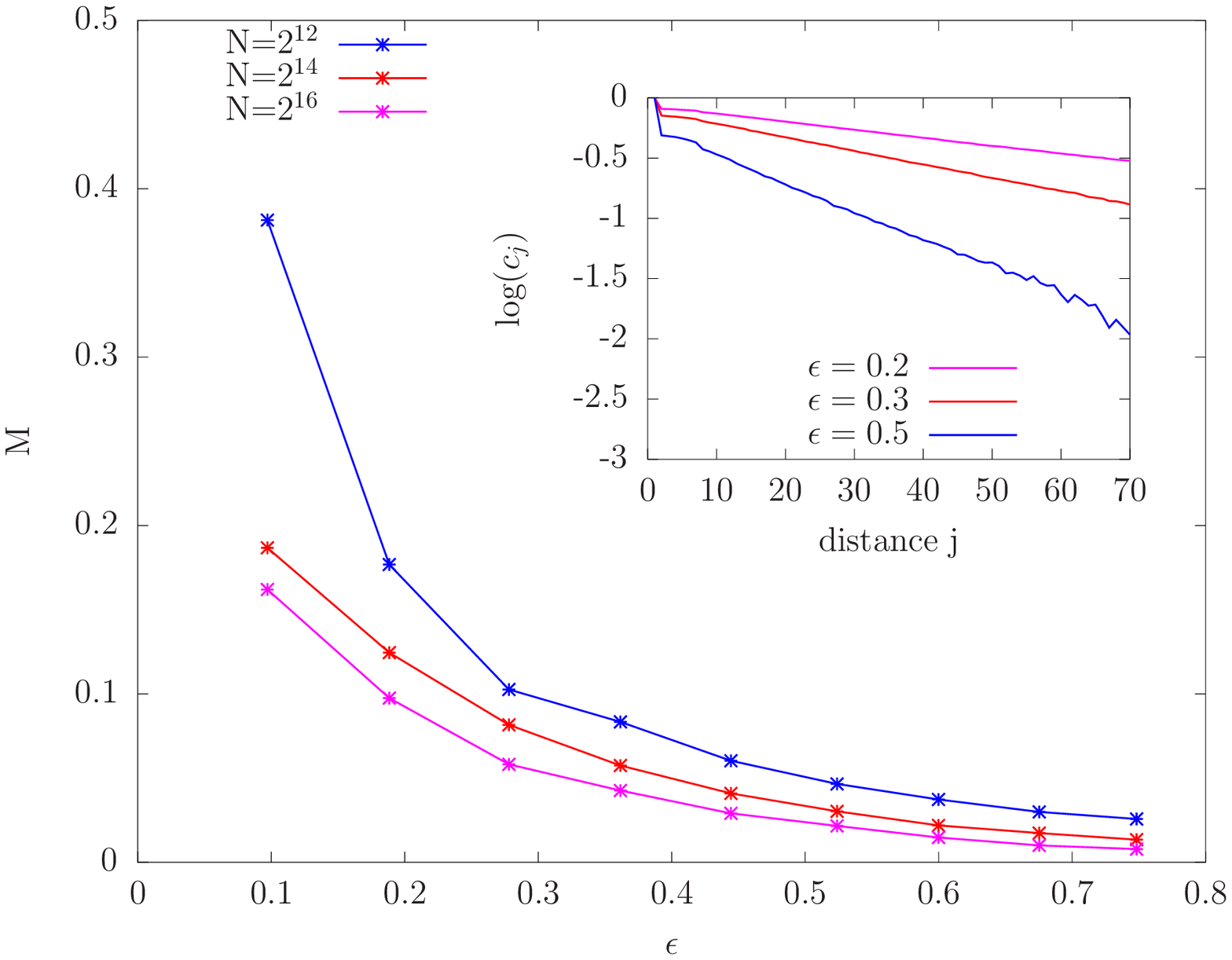}
\par\end{centering}
\begin{centering}
(b)\includegraphics[width=8cm]{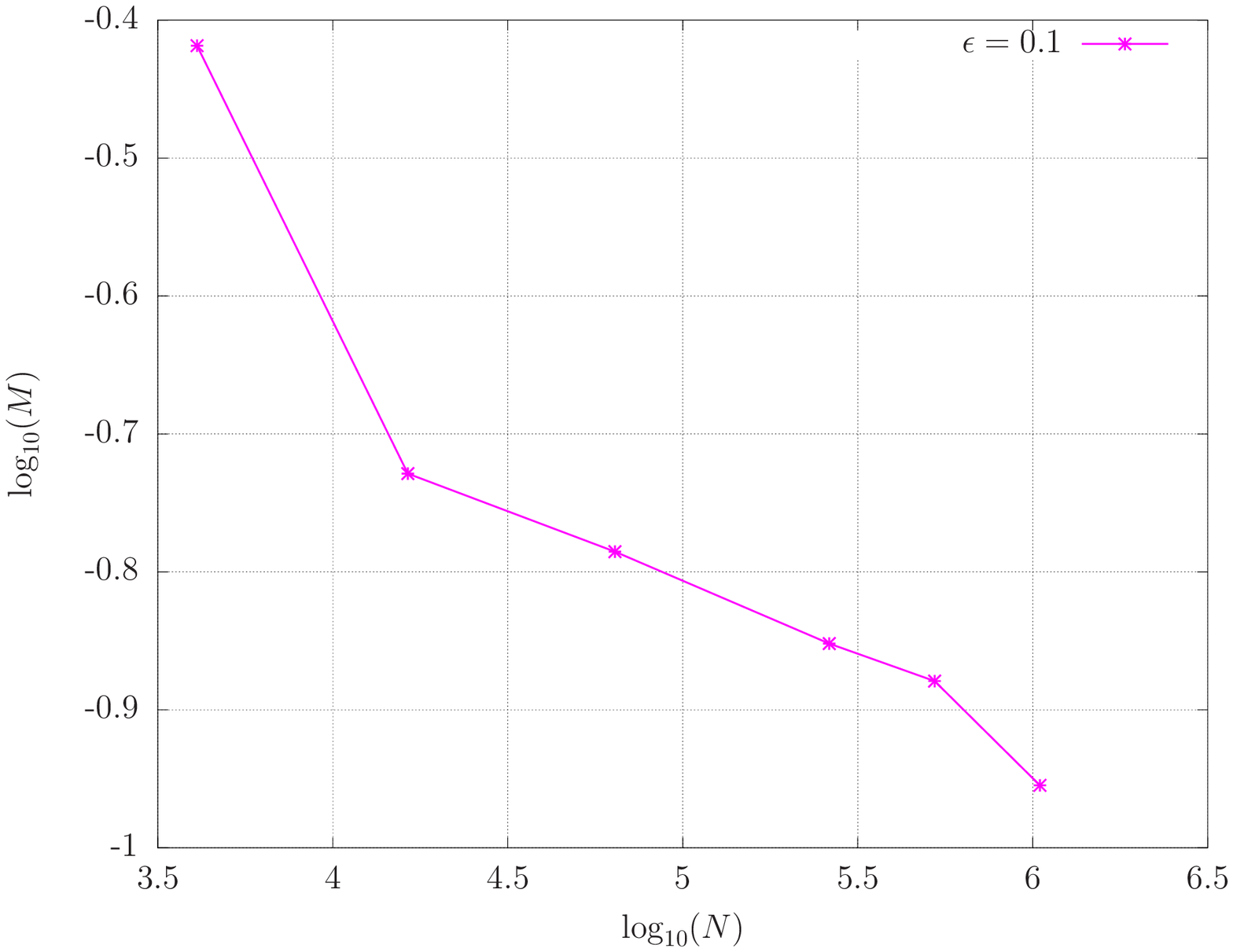}

\par\end{centering}

\caption{(colour online) (a) Equilibrium magnetisation versus the energy density $\epsilon=E/N$ for 
$\gamma=1.25$ and different sizes. {The errorbars are of the size of the dots}; (inset) Correlation function $c_{j}$ for $\gamma=1.25$ and $N=2^{14}$. {(b) Residual magnetisation for $\gamma=1.25$ at $\epsilon=0.1$ versus the system size. The simulations up to $N=2^{16}$ have a duration 
of $T_{f}=20000$, while for $N>2^{18}$ we set $T_{f}=30000$. We took the temporal mean on the second half of the simulation, after having checked the reaching of the equilibrium.} }
\label{Flo:magnetisation 1.25}
\end{figure}
to test the eventual presence of a Kosterlitz-Thouless transition,
we monitored the correlation function:
\begin{equation}
c(j)=\frac{1}{N}\sum_{i=1}^{N}\cos(\theta_{i}-\theta_{i+j[N]}).\label{eq:correlation}
\end{equation}
\\
At equilibrium, the correlation decays exponentially (See inset in
Fig.~\ref{Flo:magnetisation 1.25}) at any {$\epsilon$} in the considered
physical range, confirming the absence of the aforementioned phase
transition.
 For {$\gamma<1.5$}, we  conclude that the
number of links is still too low to entail a change in the $1$-D
behaviour. It is interesting to notice that even a configuration
with quite a large neighborhood per spin like $\gamma=1.4$ corresponds to short range interactions.

Symmetrically, the other important range to consider is $\gamma>1.5$,
when we approach the full coupling of the spins. As shown in Fig.~\ref{fig:(up)-Time-series}a {for $\gamma=1.75$},
the mean field transition of the order parameter is recovered in this regime: it is worth stressing here 
that we recover the meanfield result even for $\gamma$ significantly lower than 2, e.g. for
$\gamma=1.6$, implying that global coherence is still reachable with
a weaker condition than the full coupling. Naturally, in Fig.~\ref{fig:(up)-Time-series}a,
a shift exists between the simulations {at $\gamma=1.75$}, performed at finite size,
and the theoretical curve which {is obtained in the mean field case and in the thermodynamic limit}. Nevertheless this interval shrinks with increasing size and it is a finite size artefact. 
{In order to check the convergence towards equilibrium and the influence of finite size effects for both regimes, $\gamma<1.5$ and $\gamma>1.5$ , we monitored the variance of the
magnetisation $\sigma^{2}=\overline{(M-\overline{M})^{2}}$ and verified that it is inversely proportional to the system size  $N$ and thus vanishes in the thermodynamic limit.}

The transition between the $1$-D behaviour and the mean field phase
appears to be critical for $\gamma_{c}=1.5$.  {As illustrated in Fig.~\ref{fig:(up)-Time-series}a, the
 magnetisation curve for $\gamma_{c}$ never recoups the mean-field one even in the thermodynamic limit. Moreover} for low energies $0.3\leq\epsilon\leq0.75$
the magnetisation is affected by important fluctuations  (Fig.~\ref{fig:(up)-Time-series}b) and it is
not clear if {the equilibrium state
\begin{figure}
\begin{centering}
(a)\includegraphics[width=8cm]{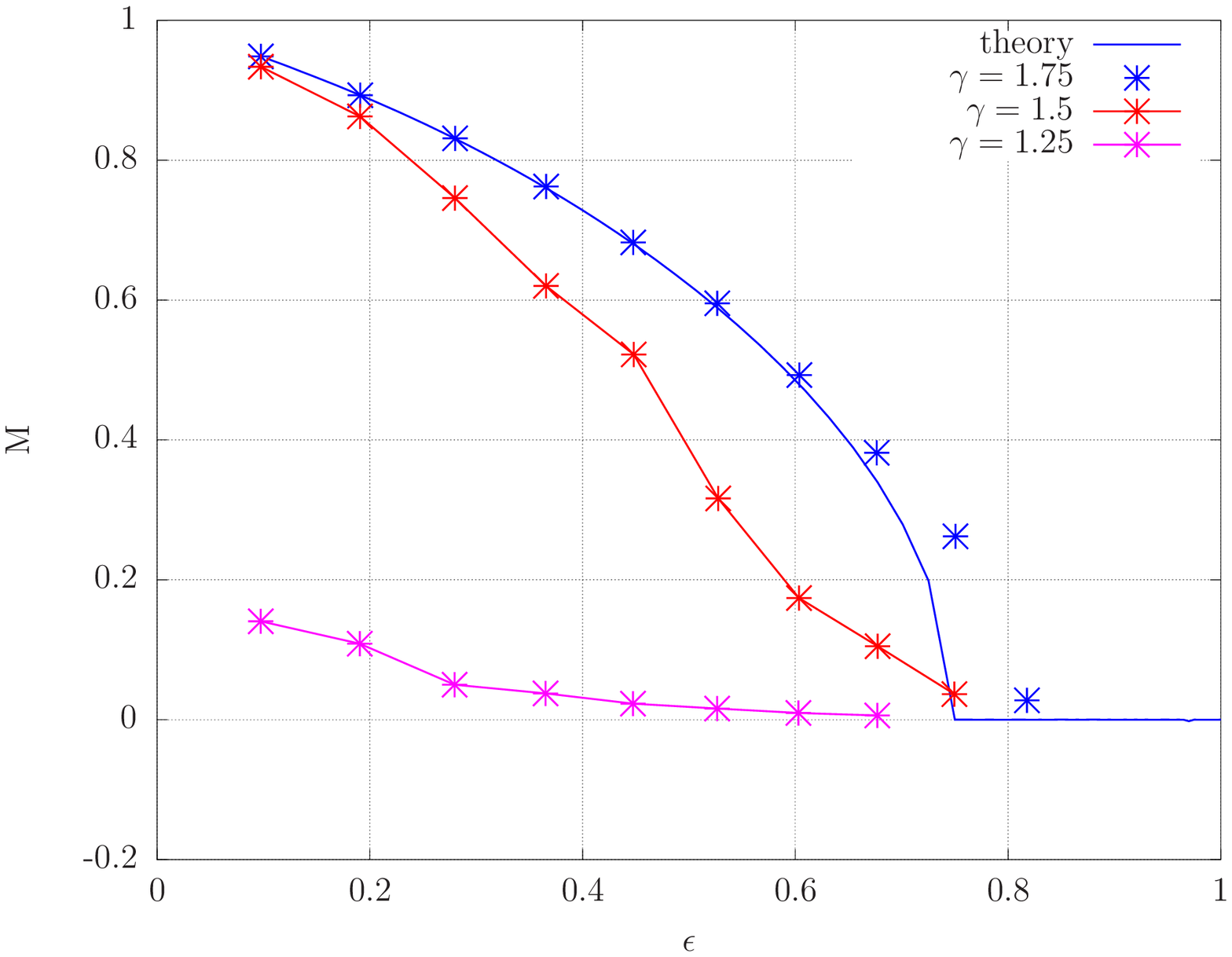}
\par\end{centering}

\begin{centering}
(b)\includegraphics[width=8cm]{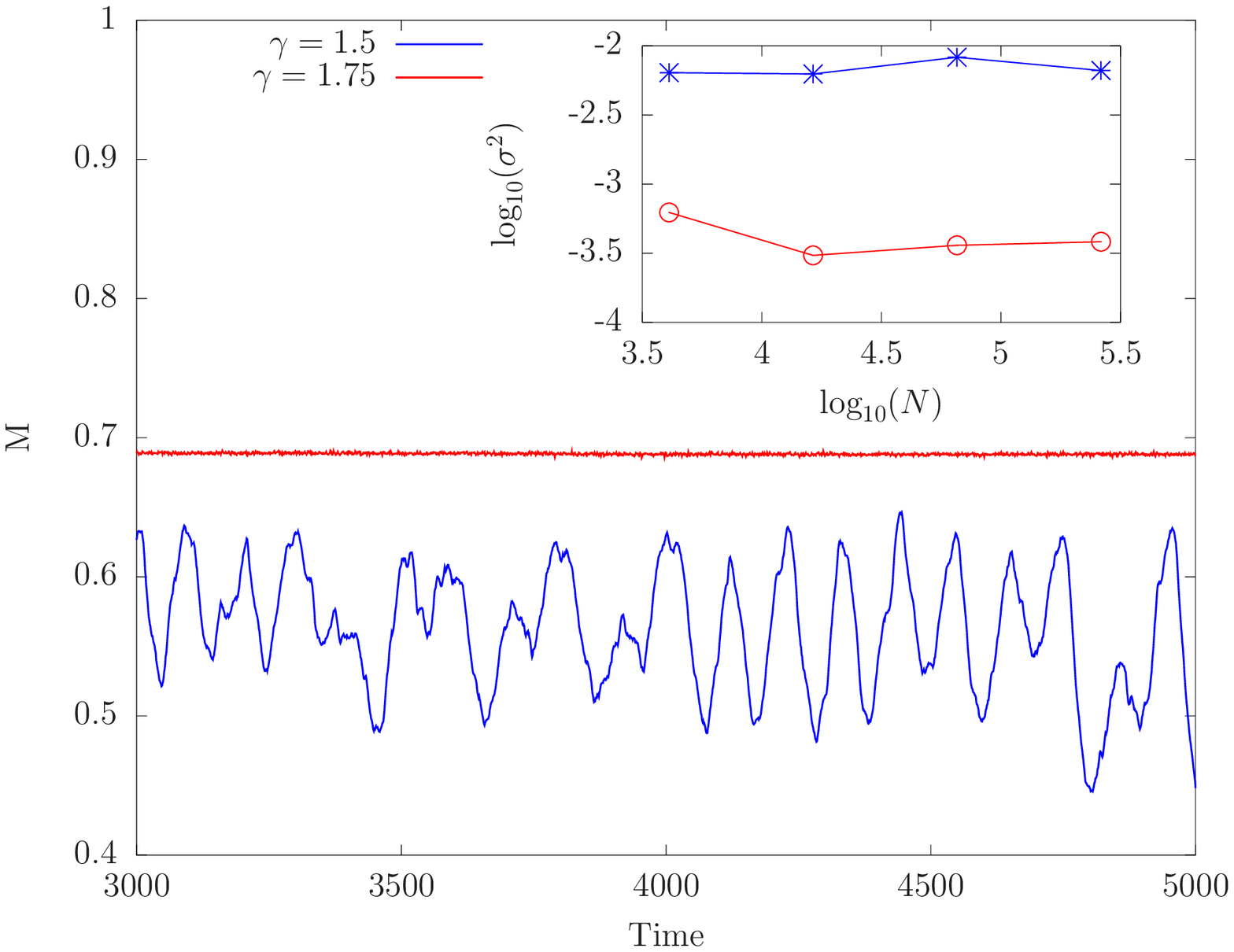}
\par\end{centering}

\caption{(colour online) (a) Equilibrium magnetisation for $N=2^{16}$ and different
$\gamma$. {For $\gamma\ne1.5$ the error bars are of the size of the dots}. (b) Time series for the
order parameter with $N=2^{18}$  and $\epsilon=0.6$; (inset) Scaling of the magnetisation variance $\left\langle \sigma^{2}\right\rangle$ for $\gamma=1.5$, $\epsilon=0.60$ (stars) and $\epsilon=0.74$
(dots).\label{fig:(up)-Time-series}}
\end{figure}
simply does not exist or it is just not reached on the time scales considered. But most likely these fluctuations are the result of the critical behaviour at $\gamma=\gamma_c$ and will persist forever, 
reflecting the ``hesitations'' of the system to reach the mean field magnetised state or the disordered low-range one.} In addition to this,
the correlation function in Eq.~(\ref{eq:correlation}) does not
prove helpful in characterising this peculiar state: it acquires the
exponential behaviour only for densities of energy above $\epsilon=0.7$,
while in the interesting interval of energies it is heavily affected
by the fluctuations and it is impossible to properly determine its
behaviour. We observed these effects on several sizes from $N=2^{12}$
up to $N=2^{18}$ and, when considering the scaling of $\sigma^{2}$
with the size (reported in the inset in Fig.~\ref{fig:(up)-Time-series}b),
it {appears} that the variance is not affected by  increasing
system size. {This phenomenon is quite peculiar as it does not occur for $\gamma>1.5$ and $\gamma<1.5$. 
Moreover these persistent fluctuations of the magnetisation tend to suggest that in this phase the system has actually an infinite susceptibility $\chi$ when it is defined as 
\begin{equation}
\chi\sim\lim_{N\rightarrow\infty}N\sigma^{2}\: .\label{chi}
\end{equation}}

We {now} argue that { $k=\sqrt{N}$, which corresponds to $\gamma=1.5$, is the lowest value of connections per spin to allow the rise of long range order. Hence, to shed light on
the mechanism underneath this topological transition}, we derive a {low energy} analytical form for the magnetisation
which shows that the critical factor is embedded in the {spectrum of the adiacency
matrix.} As {mentioned}, our first hypothesis is that we restrict our analysis
to the low {energy} regime, {which corresponds to the magnetised phase for $\gamma>1.5$}. {Having the mean field picture in mind with a magnetisation close to $1$ and considering for instance $Q=0$, 
we can assume that most spins will not deviate much from the direction of the magnetisation, which itself does not fluctuate much. 
We shall hence assume} that the difference {
$\theta_{i}-\theta_{j}$ is small for connected rotors (when $\epsilon_{i,j}=1$) \cite{Neglected_modulos}}.
 We therefore obtain a simple quadratic Hamiltonian:
$H=\sum_{i}\frac{p_{i}^{2}}{2}+\frac{J}{4k}\sum_{i,j}\epsilon_{ij}(\theta_{i}-\theta_{j})^{2}.$

To proceed further, we consider a representation for the $\{\theta_{i},p_{i}\}$
as a sum of random phased {modes} \cite{Leoncini2001, boundary_conditions}:
\begin{equation}
\begin{array}{c}
\theta_{i}=\sum_{l}\alpha_{l}(t)\cos(\frac{2\pi li}{N}+\phi_{l}),\\
p_{i}=\sum_{l}\dot{\alpha}_{l}(t)\cos(\frac{2\pi li}{N}+\phi_{l}),
\end{array}\label{eq:representation}
\end{equation}
where $\phi_{l}$ are randomly distributed phases on the circle. Since
we make the hypothesis that the time dependence is totally encoded
in the amplitudes $\alpha_{l}$, the momenta are simply related to
the angles via the first Hamilton equation $\dot{\theta_{i}}=p_{i}$.
The basic idea behind this reasoning is that, at equilibrium, the
momenta are Gaussian distributed variables, justifying the representation
in Eqs. (\ref{eq:representation}). We also observe that it consists
in a linear changing of variable since we use $N$ modes for our representation.
If we now consider different sets of phases $\{\phi_{l}\}_{m}$ labeled
by $m$, we have that each one of them corresponds to a phase space
trajectory and, hence, it is possible to replace the ensemble average
with the average on the random phases \cite{Leoncini2001}. Consequently,
injecting Eqs. (\ref{eq:representation}) in the linearised Hamiltonian
and averaging on the random phases, we obtain:
\begin{equation}
\frac{\left\langle H\right\rangle }{N}=\frac{1}{2}\sum_{l=1}^{N}\dot{\alpha}_{l}^{2}+\alpha_{l}^{2}(1-\lambda_{l}),\label{alpha hamiltonian}
\end{equation}
where 
\begin{equation}
\lambda_{l}=\frac{2}{k}\sum_{m=1}^{k/2}\cos(\frac{2\pi ml}{N})\label{eq:eigen}
\end{equation}
are the eigenvalues of the adiacency matrix. Using the second Hamilton
equation $\frac{d}{dt}(\frac{\partial\left\langle H\right\rangle }{\partial\dot{\alpha_{l}}})=-\frac{\partial\left\langle H\right\rangle }{\partial\alpha_{l}}$,
we obtain from Eq. (\ref{alpha hamiltonian}) a dispersion
relation of the wave amplitudes that embeds two levels of information:
at the microscopic level, the structure of the links, via the adiacency
matrix spectrum and, from a more macroscopical point of view, Eq.
(\ref{alpha hamiltonian}) results from averaging on the random phases
which, as explained, accounts for the ensemble averaging.
 Imposing the equipartition of {the kinetic} energy at equilibrium for the obtained collection
of harmonic oscillators (see \cite{Leoncini2001})  gives an additional relation between the
frequencies $\omega_{l}$ and the amplitudes $\alpha_{l}$: $\alpha_{l}^{2}\omega_{l}^{2}=2T/N${,
where $T$ is the kinetic temperature}
We evaluate now the magnetisation in the low temperature regime using
the same approach: we inject the representation (\ref{eq:representation})
and we average on the phases, obtaining \cite{Leoncini98}:\textbf{
\begin{equation}
\left\langle \mathbf{M}\right\rangle =\prod_{l}J_{0}(\alpha_{l})(\cos\theta_{0},\sin\theta_{0}),\label{eq:prod bessels}
\end{equation}
}where $\theta_{0}$ is the average of the $\{\theta_{i}\}$ which
is a constant because of the conservation of the total momentum $P=0$.
The absolute value of the magnetisation $\left\langle M\right\rangle $
will hence be, from Eq.~(\ref{eq:prod bessels}), the product over
the $l$ modes of the Bessel functions. To evaluate the logarithm
of $\left\langle M\right\rangle $ , we observe that, at equilibrium
and in the limit of large system size, we expect to have small $\alpha_{l}^{2}$.
We can thus approximate the Bessel functions in the limit of small
amplitudes $\alpha_{l}$ which is, therefore, the low temperatures
regime. This finally leads to:
\begin{equation}
\ln(\left\langle M\right\rangle )=-\sum_{l}\frac{\alpha_{l}^{2}}{4}=-\frac{T}{2N}\sum_{l}\frac{1}{1-\lambda_{l}}.\label{eq:magnetisation final}
\end{equation}
We calculated numerically Eq. (\ref{eq:magnetisation final}) for
increasing N and in Fig. \ref{fig:Approximated-magnetisation-} we
show how it 
\begin{figure}
\centering{}
\includegraphics[width=8cm]{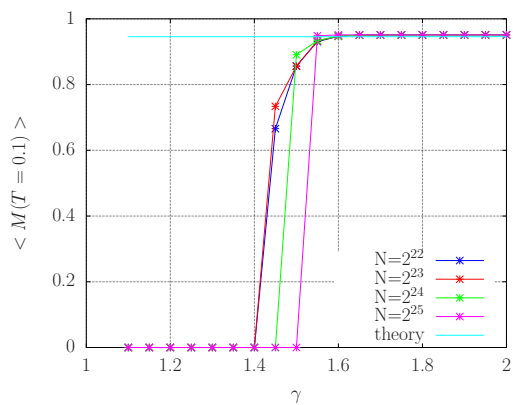}
\caption{(colour online) Approximated magnetisation $\left\langle M\right\rangle$ from Eq.(\ref{eq:magnetisation final}) for $T=0.1$
 versus  $\gamma$.{ Theory refers to the theoretical value obtained in the mean field situation.}}
\label{fig:Approximated-magnetisation-}
\end{figure}
 correctly {captures} the behaviour of the magnetisation: in the low
temperature regime, it retrieves the {mean field} value for $\gamma>1.5$
and it vanishes when $\gamma<1.5$. Moreover, with increasing
size, the difference between the two regimes becomes sharper confirming
the critical nature of $\gamma_{c}=1.5$. The key for this peculiar
effect at $\gamma=1.5$ appears thus to be fully encoded in the spectrum
of the adiacency matrix, which drives the system to the mean field
regime or to the short range one according to 
$\gamma$. Nevertheless, by a rapid inspection of Eq. (\ref{eq:eigen}),
it appears non trivial to isolate the dependence of the eigenvalues
on {$\gamma$} and on the size {since} each eigenvalue consists of a sum
of $k/2$ contributions.
\begin{figure}
\includegraphics[width=8cm]{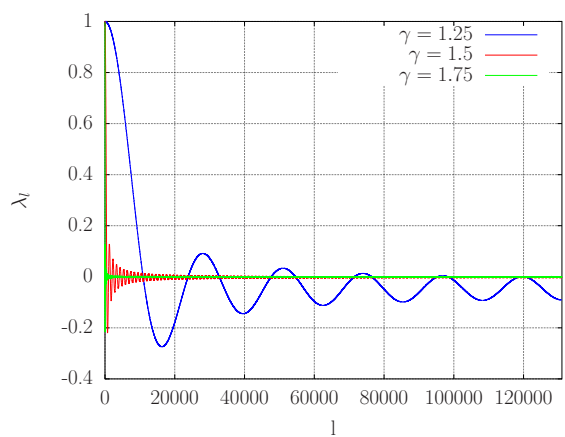}
\caption{(colour online) Spectra $\lambda_{l}$ for $N=2^{18}$ and different $\gamma$ values.\label{fig:Spectra}}
\end{figure}
 In Fig. \ref{fig:Spectra}, we show the behaviour of the spectrum
for three representative values of $\gamma$: clearly the spectra
qualitatively differ, but how to quantify
this difference is still object of a more refined analysis to precisely
relate the spectrum properties to the magnetisation behaviour.

In this Letter we 
 introduce {a} model for the interaction,
 and focused on the regular lattice topology in which
{we monitored the length of the interaction by  controlling} the degree of each spin via the parameter $\gamma$.
We showed that three different regimes existed:{ the interval $\gamma<1.5$}, where  long-range order is absent,
{ a highly connected phase ($\gamma>1.5$)} in which the 
mean field behaviour is recovered and a peculiar behaviour at the threshold of $\gamma=1.5$.
Interestingly, we show that the mean field transition does not {require}
the full coupling of the spins, like in the HMF model or in a random network \cite{diluted_networks}, and it still arises for a regular topology even for
$\gamma=1.6$ , quite far hence from the extremal configuration of
$\gamma=2$. 
However, the main result of our analysis
is the evidence of a unsteady almost turbulent like state when $\gamma=1.5$:
 the important fluctuations affecting the order parameter and the
invariance of these effects on the system size in a whole interval
of energies are in total contrast with what observed in the other
regimes, where with the same initial conditions the convergence to
equilibrium is rapid. We present a analytical calculation for the
magnetisation 
 which is able to catch the appropriate behaviour in the two limits discussed
before. This result points out that $\gamma=1.5$ is indeed the critical
value for this passage from the $1$-D topology to the mean field
frame. Moreover, it {shows} that the spectrum of the adiacency matrix,
which carries the information on the links, is crucial to understand
this shift. 
{Hence} this unstable state stems from
\emph{topological features} of the lattice, instead of from a particular
choice of the initial conditions as in \cite{Antoniazzi2007,ruffo2007,maximum_entropy}.
We anticipate that the same kind of ``bifurcation`` phenomenon could be observed with different topologies
and probably lower {connectivities}. We also believe that if we were able to find an
efficient way to modify the  parameter {$\gamma$}, these systems could
prove to be useful {adaptable} on-off switches for a somewhat larger {energy/}temperature
range{, as adding or removing a few links totally changes the macroscopic behaviour.}

\acknowledgments
The authors are grateful to W. Ettoumi for discussions. S.d.N. has
been supported by DGA/DS/MRIS.

\bibliographystyle{eplbib}

\begin{thebibliography}{10}
\bibitem{Neglected_modulos}
The approximation made corresponds to neglecting the the modulo in $\delta\theta\approx 0 [2\pi]$. 
In the low temperature regime, the width of the angle distribution is so small and, by our choice, centered on $0$ so that this appears as a legitimate assumption.
\bibitem{boundary_conditions}
It is important to note that we considered periodic boundary conditions, in our simulations and our analytical treatment
and to keep in mind that these conditions  may influence the results once the long range effects are at play.
\end{thebibliography}

\end{document}